\documentclass[showpacs,aps,graphicx,twocolumn]{revtex4}
\usepackage{graphicx}

\begin{document}

\title{Deterministic Secure Quantum Communication Without Maximally Entangled States}
\author{ Xi-Han Li, Fu-Guo Deng\footnote{Email address: fgdeng@bnu.edu.cn},
 Chun-Yan Li, Yu-Jie Liang, Ping Zhou and Hong-Yu Zhou }
\address{ The Key Laboratory of Beam Technology and Material Modification of Ministry of Education,
Beijing Normal University, Beijing 100875, People's Republic of
China, and\\ Institute of Low Energy Nuclear Physics, and Department
of Material Science and Engineering, Beijing Normal
University, Beijing 100875, People's Republic of China, and\\
Beijing Radiation Center, Beijing 100875, People's Republic of
China}
\date{\today }

\begin{abstract}
Two deterministic secure quantum communication schemes are proposed,
one based on pure entangled states and the other on $d$-dimensional
single-photon states. In these two schemes, only single-photon
measurements are required for the two authorized users, which makes
the schemes more convenient than others in  practical applications.
Although each qubit can be read out after a transmission of
additional classical bit, it is unnecessary for the users to
transmit qubits double the distance between the sender and the
receiver, which will increase their bit rate and their security. The
parties use decoy photons to check eavesdropping efficiently. The
obvious advantage in the first scheme is that the pure entangled
source is feasible with present techniques.

\end{abstract}

 \maketitle

\section{Introduction}

In the last decade, scientists have made  dramatic progress in the
field of quantum communication \cite{book,gisin}. The quantum key
distribution (QKD), whose task is to create a private key between
two remote authorized users, is one of the most remarkable
applications of quantum mechanics. By far, there has been much
attention focused on the QKD
\cite{bb84,ekert91,bbm92,gisin,longqkd,CORE,BidQKD,ABC,guoQKD,delay}
since Bennett and Brassard (BB84) \cite{bb84} proposed an original
protocol in 1984. In recent years, a novel concept, quantum secure
direct communication (QSDC) was put forward and studied by some
groups \cite{two-step,QOTP,Wangc,bf,cai,caiA}. It allows two remote
parties to communicate directly without creating a private key in
advance and  then using it to encrypt the secret message
\cite{two-step,QOTP,Wangc,bf,cai,caiA}. Thus, the sender should
confirm whether the channel is secure before he encodes his message
on the quantum states because the message cannot be discarded,
unlike that in QKD protocols \cite{two-step,QOTP,Wangc}. In 2002,
following some ideas in quantum dense coding \cite{bw}, Bostr\"{o}m
and Felbinger \cite{bf} proposed a ping-pong QSDC scheme by using
Einstein-Podolsky-Rosen (EPR) pairs as quantum information carriers,
but it has been proven to be insecure in a noise channel
\cite{attack}. Recently, Deng \emph{et al.} \cite{two-step} proposed
a two-step QSDC scheme with an EPR pair block and another scheme
with a sequence of single photons \cite{QOTP}. Wang \emph{et al.}
\cite{Wangc} introduced a high-dimensional QSDC protocol by
following some ideas in quantum superdense coding \cite{superdense}.
Now, QSDC has also been  studied in the case of a network
\cite{Linetwork,dengnetwork,dengepl}.

Another class of quantum communication protocols
\cite{imo1,beige,zhangzj,yan,Gao,zhangs2006,wangj,song,leepra} used
to transmit secret messages is called deterministic secure quantum
communication (DSQC). Certainly, the receiver can read out the
secret message only after he exchanges at least one bit of classical
information for each qubit with the sender in a DSQC protocol, which
is different from QSDC. DSQC is similar to QKD, but it can be used
to obtain deterministic information, not a random binary string,
which is different from the QKD protocols
\cite{bb84,ekert91,bbm92,gisin} in which the users cannot predict
whether an instance is useful or not. For transmitting a secret
message, those protocols
\cite{imo1,beige,zhangzj,yan,Gao,zhangs2006,wangj,song,leepra} can
be replaced with an efficient QKD protocol, such as those in Refs.
6-11, because the users can retain or flip the bit value in the key
according to the secret message after they obtain the private key
\cite{QOTP}. Schimizu and Imoto \cite{imo1} and Beige \emph{et al.}
\cite{beige} presented some novel DSQC protocols with entanglement
or a single photon. More recently, Gao and Yan \cite{yan, Gao} and
Man \emph{et al.} \cite{zhangzj} proposed several DSQC schemes based
on  quantum teleportation \cite{teleportation} and entanglement
swapping \cite{entanglementswapping}. The users should complete the
eavesdropping check before they take a  swapping or teleportation.
Although the secret message can be read out only after transmitting
an additional classical bit for each qubit, do not have the users to
transmit the qubits that carry the secret message. Therefore, these
schemes may be more secure than others in a noise channel, and they
are more convenient for quantum error correction. On the other hand,
a Bell-basis measurement is required inevitably for the parties in
both entanglement swapping \cite{entanglementswapping} and quantum
teleportation \cite{teleportation}, which will increase the
difficulty of implementing these schemes in laboratory.

In Ref. 35, Yan and Gao introduced an interesting DSQC protocol
following some ideas in Ref. 11 with EPR pairs. After sharing a
sequence of EPR pairs securely, the two parties of a quantum
communication only need perform single-photon measurements on their
photons and can communicate directly by exchanging a bit of
classical information for each qubit. Obviously, their DSQC protocol
is more convenient than other quantum communication protocols
\cite{bf,two-step,Wangc,zhangzj,yan,Gao,zhangs2006,caiA,song} from
the aspect of measurement even though it requires the two parties to
exchange classical bit  and each EPR pair can only carry one bit of
the message.

In this paper, we will first propose a new DSQC scheme with pure
entangled states, nonmaximally entangled two-photon states. The
quantum signal source is in a more general formal of entanglement,
which makes this scheme more suitable for applications than the
Yan-Gao protocol \cite{yandelay}. Then, we will discuss it with a
sequence of $d$-dimensional single photons. We use some decoy
photons to ensure the security of the whole quantum communication.
In both schemes, single-photon measurements are enough. Moreover, we
redefine the total efficiency of quantum communication. Compared
with the old one presented in Ref. 36, our definition is more
reasonable.

\section{DSQC with pure entangled states}

\subsection{DSQC with Two-dimensional Quantum Systems}

In the DSQC schemes with entanglement swapping and teleportation
\cite{yan,Gao,zhangzj}, the parties usually use EPR pairs as the
quantum information carriers. An EPR pair is in one of the four Bell
states, the four maximally two-qubit entangled states, as follows:
\begin{eqnarray}
\vert \psi^{\pm} \rangle_{AB}=\frac{1}{\sqrt{2}}(\vert 0 \rangle_A
\vert 1 \rangle_B \pm \vert 1 \rangle_A \vert 0 \rangle_B),\\
\vert \phi^{\pm} \rangle_{AB}=\frac{1}{\sqrt{2}}(\vert 0 \rangle_A
\vert 0 \rangle_B \pm \vert 1 \rangle_A \vert 1 \rangle_B),
\end{eqnarray}
where $\vert 0 \rangle$ and $\vert 1 \rangle$ are the eigenvectors
of the measuring basis (MB) $Z$. The subscripts $A$ and $B$ indicate
the two correlated photons in each EPR pair. For the Bell state
$\vert \psi^{\pm} \rangle$ ($\vert \phi^{\pm} \rangle$), if the two
photons are measured with the same MB $Z$, the outcomes will always
be anti-correlated (correlated). The correlation of the entangled
quantum system plays an important role in quantum communication
\cite{ekert91,bbm92,longqkd,CORE} as it provides a tool for checking
eavesdropping. For example, the two photons $A$ and $B$ are
anti-correlated in the Bennett-Brassard-Mermin 1992 QKD protocol
\cite{bbm92} even though the users measure them with the MB $Z$ or
$X$ as
\begin{eqnarray}
\vert \psi^{-} \rangle_{AB} &=& \frac{1}{\sqrt{2}}(\vert 0 \rangle_A
\vert 1 \rangle_B - \vert 1 \rangle_A \vert 0 \rangle_B)\nonumber\\
&=& \frac{1}{\sqrt{2}}(\vert +x \rangle_A \vert -x \rangle_B - \vert
-x \rangle_A \vert +x \rangle_B).
\end{eqnarray}
Here, $\vert \pm x\rangle=\frac{1}{\sqrt{2}}(\vert 0\rangle \pm
\vert 1\rangle)$ are the two eigenvectors of the basis $X$. This
nature forbids an eavesdropper to eavesdrop on the quantum
communication  freely with an intercepting-resending strategy.

In experiment, however, the two photons are usually not in the
maximally entangled state $\vert \psi^-\rangle_{AB}$. That is, a
practical quantum signal source usually produces a pure entangled
state, such as $\vert \Psi\rangle_{AB}=a \vert 0\rangle_{A}\vert
1\rangle_{B} +b\vert 1\rangle_{A} \vert 0\rangle_{B}$ (here $\vert
a\vert ^2 + \vert b\vert ^2 =1$). In this time, the two photons are
always anti-correlated with the basis $Z$, but not with the basis
$X$, as
\begin{eqnarray}
\vert \Psi\rangle_{AB} &=& a \vert 0\rangle_{A}\vert 1\rangle_{B}
+b\vert 1\rangle_{A} \vert 0\rangle_{B}\nonumber\\
&=& \frac{1}{2}[(a+b)(\vert +x\rangle_A\vert +x\rangle_B - \vert
-x\rangle_A\vert -x\rangle_B) \nonumber\\
&-& (a-b)(\vert
+x\rangle_A\vert -x\rangle_B - \vert -x\rangle_A\vert +x\rangle_B)].
\end{eqnarray}
That is, the security of the quantum communication with pure
entangled states is lower than that with Bell states if the users
use the two bases $Z$ and $X$ to measure them for the eavesdropping
check directly. On the other hand, the quantum source is more
convenient than maximally entangled states.

In the present DSQC scheme,  we will use  pure entangled states as
the quantum information carriers for DSQC. This scheme has the
advantage of a practical entangled source and of high security with
decoy photons, compared with those in Refs. 26-28 and 35. For the
integrality of our point-to-point DSQC scheme, we give all the steps
as follows:

(1) The sender Alice prepares  $N$ two-photon ordered pairs in which
each is randomly in one of the two pure entangled states $\{\vert
\Psi \rangle_{AB},\vert \Psi' \rangle_{AB} \}$. Here, $\vert \Psi'
\rangle_{AB}=a\vert 1 \rangle_A \vert 0\rangle_B + b\vert 0
\rangle_A \vert 1 \rangle_B)$ which can be prepared by flipping the
bit value of the two photons in the state $\vert \Psi \rangle_{AB}$,
i.e, $(\sigma^{A}_x\otimes \sigma^{B}_x)\vert \Psi
\rangle_{AB}=\vert \Psi' \rangle_{AB}$, similar to Ref. 10. Alice
picks out photon $A$ from each pair to form an ordered sequence
$S_A$, say [$A_1,A_2,...A_N$], and the other partner photons compose
the sequence $S_B$ =[$B_1,B_2,...B_N$], similar to Refs. 6, 13, 37,
38.

For checking eavesdropping efficiently, Alice replaces some photons
in the sequence $S_B$ with her decoy photons $S_{de}$, which are
randomly in one of the states $\{\vert 0\rangle, \vert 1\rangle,
\vert +x\rangle, \vert -x\rangle\}$. They can be prepared with an
ideal single-photon source. Also, Alice can get a decoy photon by
measuring  photon $A$ in a photon pair $\vert \Psi \rangle_{AB}$ in
the sequence $S_A$ with the MB $Z$ and then operating on photon $B$
with the local unitary operation $\sigma_x$ or a Hadamard (H)
operation:
\begin{eqnarray}
H\vert 0\rangle =\vert +x\rangle, \;\;\;\; H\vert 1\rangle =\vert
-x\rangle.
\end{eqnarray}
We will discuss the reason that Alice inserts the decoy photons in
the sequence $S_B$ in detail below.

(2) Alice encodes her secret message $M_A$ on the photons in the
sequence $S_B$ with the two unitary operations $I$ and $U=\sigma_x$,
which represent bits 0 and 1, respectively. Obviously, she can
choose all the decoy photons, $S_{de}$, as samples  for checking
eavesdropping.

(3) Alice sends sequence $S_B$  to Bob and always keeps the sequence
$S_A$ at home.

(4) After Bob confirms the receipt of sequence $S_B$, Alice tells
Bob the positions and the states of the decoy photons $S_{de}$. Bob
performs a suitable measurement on each photon in $S_{de}$ with the
same basis as Alice chose for preparing it, and completes the error
rate analysis of the samples. If the error is very low, Alice and
Bob continue their communication to next step; otherwise, they
abandon the result of the transmission and repeat the quantum
communication from the beginning.

(5) Alice and Bob measure their photons remaining in the sequences
$S_A$ and $S_B$ with the same basis $Z$, and they get the results
$R_A$ and $R_B$, respectively.

(6) Alice publicly broadcasts her results $R_A$.

(7) Bob reads out the secret message $M_A$ with his outcomes $R_B$
directly; i.e., $M_A=R_A\oplus R_B\oplus 1$.

It is interesting to point out that it is unnecessary for the
receiver Bob and the sender Alice to perform Bell-basis measurements
on their photons, they only perform single-photon measurements,
which makes this scheme more convenient than others
\cite{yan,Gao,zhangzj} with entanglement swapping and quantum
teleportation. Moreover, the quantum sources are just a practical
pure entangled states, not maximally entangled states, which makes
this DSQC scheme easier than those with Bell states
\cite{two-step,yandelay}. As each photon just transmits the distance
between the sender and the receiver, its bit rate is higher than
those with two-way quantum communication as the attenuation of a
signal in a practical channel is exponential; i.e.,
$N_s(L)=N_s(0)e^{-\lambda L}$. Here, $N_s(L)$ is the photon number
after being transmitted the distance $L$, and $\lambda$ is the
attenuation parameter.

As the security of a quantum communication scheme depends on the
error rate analysis of samples chosen randomly, the present DSQC
scheme can be made to be secure as the decoy photons are prepared
randomly in one of the four states $\{\vert 0\rangle, \vert
1\rangle, \vert +x\rangle, \vert -x\rangle\}$ and are distributed in
the sequence $S_B$ randomly. An eavesdropper, say Eve, knows neither
the states of the decoy photons nor their positions in the sequence
$S_B$, so her action will inevitably perturb the decoy photons and
be detected by the users. As the basis for the measurement on each
decoy photon is chosen after the sender has told the receiver its
basis, all of the decoy photons can be used for checking
eavesdropping, not just a fraction of them as Ref. 9.

Without the decoy photons, the security of the present DSQC scheme
will decrease as the two photons in a pure entangled state $\vert
\Psi\rangle$ or $\vert \Psi'\rangle$ have not deterministic relation
when they are measured with the MB $X$. That is, the parties cannot
determine whether the errors of their outcomes comes from
eavesdropping done by Eve or the nondeterministic relation obtained
with the MB $X$ if they only transmit a sequence of pure entangled
states. In this way, Eve can obtain a fraction of the secret message
without being detected.

\subsection{DSQC with d-dimensional Quantum Systems}

It is straightforward to general our DSQC scheme to the case with
$d$-dimensional quantum systems (such as the orbit momentum of a
photon \cite{OAM}). A pure  symmetric $d$-dimensional two-photon
entangled state can be described as
\begin{equation}
\vert \Psi_{p}\rangle_{AB} =\sum_{j} a_j\vert j\rangle_A \otimes
\vert j \rangle_B,
\end{equation}
where
\begin{equation}
\sum_{j} |a_j|^2=1.
\end{equation}
Defining
\begin{equation}
U_{m} =\sum_{j} \vert j+m\;{\rm mod} \; d \rangle \langle j\vert,
\end{equation}
which is used to transfer the state $\vert j\rangle$ into the state
$\vert j+m\rangle$; i.e.,
\begin{equation}
(U^A_{m}\otimes U^B_{m}) \vert \Psi_{p}\rangle_{AB} =\sum_{j}
a_j\vert  j+m\;{\rm mod} \; d \rangle_A \otimes \vert  j+m\;{\rm
mod} \; d  \rangle_B,\nonumber\\
\end{equation}
where $m=1,2,\cdots, d-1$.

As in Ref. 23, the MB $Z_{d}$ is made up of the $d$ eigenvectors as
\begin{eqnarray}
\left\vert  0 \right\rangle, \;\;\;\left\vert  1 \right\rangle,
\;\;\;\;\left\vert  2 \right\rangle, \;\; \cdots, \;\;\;\;\left\vert
{d - 1} \right\rangle.
\end{eqnarray}
The $d$ eigenvectors of the MB $X_{d}$ can be described as
\begin{eqnarray}
\vert 0\rangle_x&=&\frac{1}{{\sqrt d }}\left( {\left\vert  0
\right\rangle + \vert 1\rangle \;\; + \cdots \;\; + \left\vert
{d-1}\right\rangle }\right),\;\nonumber \\
\vert 1\rangle_x&=&\frac{1}{{\sqrt d }}\left({\left\vert  0
\right\rangle + e^{{\textstyle{{2\pi i} \over d}}} \left\vert  1
\right\rangle + \cdots
+ e^{{\textstyle{{(d-1)2\pi i} \over d}}} \left\vert  {d-1} \right\rangle} \right),\; \nonumber\\
\vert 2\rangle_x&=&\frac{1}{{\sqrt d }}\left({\left\vert  0
\right\rangle + e ^{{\textstyle{{4\pi i} \over d}}} \left\vert 1
\right\rangle + \cdots + e^{{\textstyle{{(d-1)4\pi i} \over d}}}
\left\vert  {d-1} \right\rangle }\right),\nonumber\\
&&\cdots \cdots \cdots \cdots \cdots \cdots \nonumber\\
\vert d-1\rangle_x&=&\frac{1}{{\sqrt d }}(\left\vert  0
\right\rangle + e ^{{\textstyle{{2(d-1)\pi i} \over d}}} \left\vert
1 \right\rangle  + e ^{{\textstyle{{2\times 2(d-1)\pi i} \over d}}}
\left\vert 2 \right\rangle + \cdots \nonumber\\
&& + e^{{\textstyle{{(d-1)\times 2(d-1)\pi i} \over d}}} \left\vert
{d-1} \right\rangle ).
\end{eqnarray}
The two vectors $\vert k\rangle$ and $\vert l\rangle_x$ coming from
two MBs satisfy the relation $\vert \langle k|l\rangle_x \vert
^2=\frac{1}{d}$. As in Ref. 23, we can construct the $d$-dimensional
Hadamard ($H_d$) operation as follows:
\begin{eqnarray}
H_d  =\frac{1}{\sqrt{d}} \left( {\begin{array}{*{20}c}
   1 & 1 &  \cdots  & 1  \\
   1 & {e^{2\pi i/d} } &  \cdots  & {e^{(d-1)2\pi i/d} }  \\
   1 & {e^{4\pi i/d} } &  \cdots  & {e^{(d-1)4\pi i/d}  }\\
    \vdots  &  \vdots  &  \cdots  & \vdots  \\
   1 & {e^{2(d-1)\pi i/d} } &  \cdots  & {e^{(d-1)2(d-1)\pi i/d} }  \\
\end{array}} \right)\label{HD}.
\end{eqnarray}
That is, $H_d\vert j\rangle=\vert j\rangle_x$.

For quantum communication, Alice prepares  $N$ ordered
$d$-dimensional two-photon pure entangled states. Each pair is
randomly in one of the states $\{ (U^A_{m}\otimes U^B_{m}) \vert
\Psi_{p}\rangle_{AB} \}$ ($m=0,1,2,\cdots, d-1$), similar to the
case with two-dimensional two-photon pure entangled states. The
uniform distribution of the pure entangled states will make the
users obtain outcomes $0$, $1$, $2$, $\cdots$, $d-1$ with the same
probability. Before the communication, Alice divides the photon
pairs into two sequences, $S_A'$ and $S_B'$. That is, the sequence
$S_A'$ is composed of photons $A$ in the  $N$ ordered photon pairs,
and the sequence $S_B'$ is made up of photons $B$.

The sender Alice can also prepare decoy photons, similar to the case
with two-dimensional photons. In detail, Alice measures some of the
photons $A$ in the sequence $S_A'$ with the basis $Z_d$ and then
operates on them with $I$ or $H_d$. She inserts the decoy photons in
the sequence $S_B'$ and keeps their positions secret. For the other
photons in the sequence $S_B'$, Alice encodes her secret message on
the sequence $S_B'$ with the unitary operations $\{U_{m}^B\}$. After
Bob receives the sequence $S_B'$, Alice requires Bob to measure the
decoy photons with the suitable bases $\{Z_d, X_d\}$, the same as
those used for preparing them. If the transmission is secure, Alice
and Bob can measure the photons remaining in the sequences $S_A'$
and $S_B'$ with the basis $Z_d$, respectively. After Alice publishes
her outcomes $R_A'$, Bob can obtain the secret message $M_A$
directly with his own outcomes $R_B'$.

\subsection{The Capacity and Efficiency}

In each pure entangled state $\rho$, such as $\vert
\Psi_{p}\rangle_{AB} =\sum_{j} a_j\vert j\rangle_A \otimes \vert j
\rangle_B$, the von Neumann entropy for each photon is
\begin{eqnarray}
S(\rho)=-\sum_i \lambda_i \; log_2 \; \lambda_i = -\sum_{i=0}^{d-1}
|a_i|^2 \; log_2 \;|a_i|^2,
\end{eqnarray}
where $\lambda_i=|a_i|^2$ is the probability that one gets the
result $\vert i\rangle$ when one measures photon $A$ or $B$ in the
state $\vert \Psi_{p}\rangle_{AB} =\sum_{j} a_j\vert j\rangle_A
\otimes \vert j \rangle_B$ with the basis $Z_d$. When
$|a_i|^2=\frac{1}{d}$ (for each $i=0,1,\cdots, d-1$), the von
Neumann entropy has its the maximal value $S(\rho)_{max}=log_2 d$.
In other cases, $S(\rho)_{max}<log_2 d$. For each photon pair, its
von Neumann entropy is $2S(\rho)$.

In fact, each pure entangled state $\rho$ in the present DSQC scheme
can carry $log_2 d$ bits of classical information. It is obvious
that photon $B$ is randomly in the state $\vert i\rangle$ with the
probability $P(i)=\frac{1}{d}$ when Bob measures it with the basis
$Z_d$. The reason is $P(i)=\frac{1}{d}\sum_m \vert a_m
\vert^2=\frac{1}{d}$ as the photon pair is randomly in one of the
states $\{(U^A_{m}\otimes U^B_{m}) \vert \Psi_{p}\rangle_{AB} \}$
($m=0,1,2,\cdots, d-1$). That is, the distribution of the pure
entangled states $\{(U^A_{m}\otimes U^B_{m}) \vert
\Psi_{p}\rangle_{AB} \}$ provides a way for carrying information
efficiently.

As almost all the quantum source (except for the decoy photons used
for eavesdropping check) can be used to carry the secret message,
the intrinsic efficiency $\eta_q$,  for qubits in our schemes
approaches 100\%. Here,
\begin{eqnarray}
\eta_q=\frac{q_u}{q_t},
\end{eqnarray}
where $q_u$ is the number of useful qubits in the quantum
communication and $q_t$ is the number of total qubits used (not the
ones transmitted; this is different from the definition proposed by
Cabello \cite{Cabello}). We define the total efficiency of a quantum
communication scheme as
\begin{eqnarray}
\eta_t=\frac{m_u}{q_t+b_t},
\end{eqnarray}
where $m_u$ and $b_t$ are the numbers of message transmitted and the
classical bits exchanged, respectively. In the present DSQC scheme,
$m_u=log_2 d$, $q_t=2S(\rho)$ and $b_t=log_2 d$ as the users pay
$\log_2 d$ bits of classical information and $q_t=2S(\rho)$ bits of
quantum information (a photon pair) for $m_u=log_2 d$ bits of the
secret message. Thus, its total efficiency is $\eta_t=\frac{log_2
d}{log_2 d + 2S(\rho)}\geq \frac{1}{3}$ in theory.

It is of interest to point out that our definition of the total
efficiency of a quantum communication scheme, $\eta_t$, is more
reasonable, compared with the old one \cite{Cabello}. Even though
Alice only transmits a sequence of photons to Bob, the source is an
entangled one, which is different from the single photons discussed
below. Obviously, the new definition can be used to distinguish a
scheme with single photons from one with entangled ones if the
efficiency for qubits and the classical information exchanged are
both the same. Moreover, the total efficiency of dense coding
according to this definition is no more than 100\% as the traveling
photon in an EPR pair carries two bits of information, and the
quantum system used for the quantum channel is a two-qubit one.

\begin{center}
\section{efficient one-way DSQC with $d$-dimensional single photons }
\end{center}

In our DSQC scheme above, the parties only exploit the correlation
of the two photons in a pure entangled state along the direction $z$
for transmitting the secret message. We can also simplify some
procedures with single photons following some ideas in Ref. 14.
Certainly, an ideal single-photon source is not available for a
practical application at present, different from the pure
entanglement source. With the development of technology, we believe
that a practical ideal single-photon source can be produced without
difficulty \cite{singlephotonsource}, so in theory, it is
interesting to study the model for DSQC with single photons.

Similar to the case with pure entangled states, we can describe the
principle of our DSQC scheme with single photons as follows:

(S1) Alice prepares a sequence of $d$-dimensional single photons
$S$. She prepares them by choosing the MB $Z_d$ or the MB $X_d$
randomly, the same as in Ref. 14. She chooses some photons as the
decoys and encodes her secret message on the other photons with the
unitary operations $\{U_m, U_m^x\}$, where
\begin{equation}
U_{m}^x =\sum_{j}e^{\frac{2\pi i}{d}jm} \vert j+m\;{\rm mod} \; d
\rangle \langle j\vert.
\end{equation}
That is, Alice encodes her secret message with the operations $U_m$
if a single photon is in one of the eigenstates of the MB $Z_d$.
Otherwise, she will encode the message with the operations $U_m^x$.

(S2) Alice sends the sequence $S$ to Bob.

(S3) Bob completes the error rate analysis on the decoy photons. In
detail, Alice tells Bob the positions and the states of the decoy
photons. Bob measures them with the suitable MBs and analyzes their
error rates.

(S4) If the transmission of the sequence $S$ is secure, Alice tells
Bob the original states of the photons retained. Bob measures them
with the same MBs as those chosen by Alice for preparing them.
Otherwise, they discard their transmission and repeat the quantum
communication from the beginning.

(S5) Bob reads out the secret message $M_A$ with his own outcomes.

In essence, this DSQC scheme is a revision of the QSDC protocol in
Ref. 14, and is modified for transmitting the secret message in
one-way quantum communication. Compared with the schemes based on
entanglement \cite{bf,two-step,Wangc,zhangzj,
yan,Gao,caiA,zhangs2006,song}, this DSQC scheme only requires the
parties to prepare and measure single photons, which makes it more
convenient in practical applications, especially with the
development of  techniques for storing quantum states
\cite{storage}. Compared with the quantum one-time pad QSDC scheme
\cite{QOTP}, the photons in the present DSQC scheme need only be
transmitted from the sender to the receiver, not double the distance
between the parties, which will increase the bit rate in a practical
channel as the channel will attenuate the signal exponentially with
the distance $L$. Certainly, the parties should exchange a classical
bit for each qubit to read out the secret message. As the process
for exchanging a classical bit is by far easier than that for a
qubit, the present scheme still has an exciting nature for
applications.

\bigskip
\section{discussion and summary}

Similar to DSQC protocols \cite{yan,Gao,zhangzj} with entanglement
swapping and teleportation, Alice can also encode her secret message
on the sequence $S_A$ after she sends the sequence $S_B$ to Bob and
confirms the security of the transmission in our first DSQC scheme.
Moreover, she can accomplish this task in a simple way. That is,
Alice first measures her photons remaining in the sequence $S_A$ and
then publishes the difference between her outcomes and her secret
message. In the DSQC scheme with single photons, Alice need only
modify the process for publishing her information to encode her
secret message after the transmission is confirmed to be secure. At
this time, she only tells Bob the combined information about the
original states of the single photons and the secret message, not
just the states.

In summary, we have proposed two DSQC protocols. One is based on a
sequence of pure entangled states, not maximally entangled ones. The
obvious advantage is that a pure-entanglement quantum signal source
is feasible at present. In the other scheme, the parties exploit
only a sequence of $d$-dimensional single photons.  In the present
two DSQC protocols, only single-photon measurements are required for
the authorized users, which makes them more convenient than those
\cite{yan,Gao,zhangzj} with quantum teleportation and entanglement
swapping. Even though it is necessary for the users to exchange one
bit of classical information for each bit of the secret message, the
qubits do not run through the quantum line twice, which will
increase their bit rate and security in practical conditions as the
qubits do not suffer from the noise and the loss aroused by the
quantum line after they are transmitted from one party to the other.
Also, they can be easily modified for encoding the secret message
after confirming the security of the quantum channel, the same as in
Refs. 26-28.

\section*{ACKNOWLEDGMENTS}

This work is supported by the National Natural Science Foundation of
China under grant Nos. 10447106, 10435020, 10254002, and A0325401
and by the Beijing Education Committee under grant No. XK100270454.

\end{document}